\documentclass[onecollarge]{svjour2}       
\smartqed  
\usepackage{psfrag,graphicx}
\journalname{Journal of Statistical Physics}
\begin{document}

\title{Strength distribution of repeatedly broken chains}

\author{Michael Wilkinson \and Bernhard Mehlig}

\institute{Michael Wilkinson \at
              Department of Mathematics, \\
              The Open University,
              Walton Hall, \\
              Milton Keynes, MK7 6AA, \\
              England.\\
              \email{m.wilkinson@open.ac.uk}      \\
           \and
           Bernhard Mehlig \at
              Department of Physics\\
              G\" oteborg University, \\
              41296 Gothenburg, \\
              Sweden.\\
              \email{mehlig@fy.chalmers.se}
}

\date{Received: date / Accepted: date}

\maketitle

\begin{abstract}
We determine the probability distribution of the breaking strength
for chains of $N$ links, which have been produced by repeatedly
breaking a very long chain. \keywords{Fragmentation \and Chain
breaking} \PACS{02.50.-r,05.40.-a}
\end{abstract}

\section{Introduction}
\label{sec: 1}

Consider a chain assembled from $N$ random links with independent,
identically distributed breaking strengths $x$, which have
probability density $\rho(x)$. It is easy to calculate the
probability density $\rho (x\vert N)$ for the strength of the
chain being less than $x$. In some contexts, however, the relevant
question is a different one: what are the strengths of chain
segments of length $N$ which are obtained by repeatedly breaking a
very long chain? Such chain segments are expected to be stronger
than their randomly assembled counterparts, because they have been
produced by a process which has eliminated the weakest links. Here
we calculate the distribution of the strength of a chain segment
of length $N$ which has been produced by repeated breaking a very
long chain, of length ${\cal N}$, say. Specifically, we break the
chain of length ${\cal N}$ at its weakest link, then break each
fragment at its weakest link and continue the process. We collect
the links of length $N$ produced by this process, and determine
the probability density of their strengths, $\rho^\ast(x\vert N)$.
In this paper we obtain formulae which are precise asymptotic
results for the limit where $N\gg 1$.

The calculation of this distribution involves the use of prior
information: we know that the chain segment of length $N$ and
breaking strength $x$ was produced by breaking a longer and weaker
chain, of length $N_0$ and breaking strength $x_0$. What makes the
problem difficult is that the prior information, in the form of
the values of $N_0$ and $x_0$, is itself uncertain. It is hard to
solve this problem for the case of a chain which has previously
been broken only once. In the case of a chain which has been
broken many times, the distributions of $x_0$ and $N_0$ themselves
depend on previous breakages. The problem appears to be very
difficult when the chain has previously been broken many times.

\begin{figure}[t]
\includegraphics[width=7cm,clip]{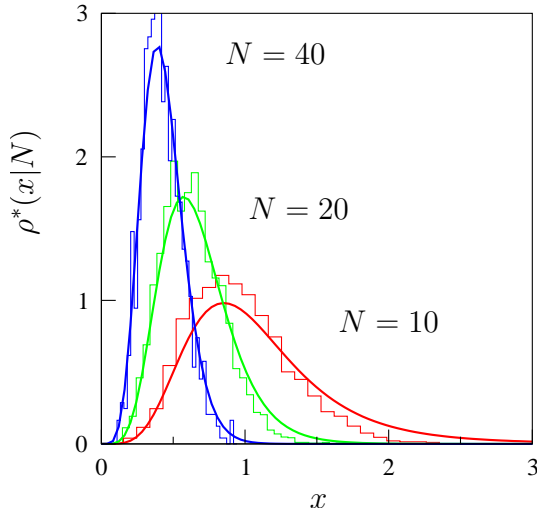}
\caption{\label{fig: 1} Shows $\rho^\ast(x\vert N)$ for three
different values of $N$, namely $10$ (red), $20$ (green)  and $40$
(blue), for $\rho(x) = x\exp(-x)$. Shown are results of computer
simulations (histogram) and those of (\ref{eq: 0}) which for the
case described here takes the form $\rho^\ast(x|N) = N^3\, x\,{\rm
e}^{-x} \{[(1+x)^2/2- (1+x){\rm e}^x] {\rm e}^{-2x}+1/2\} {\rm
e}^{-N(1-(1+x){\rm e}^{-x})}$.}
\end{figure}

We consider the case where the chain is very long, and we
anticipate that there is an asymptotic form for the strength
distribution which is independent of the initial length ${\cal
N}$. We use a self-consistent calculation and obtain
$\rho^\ast(x\vert N)$ in closed form in terms of the cumulative
strength distribution of an individual link, $P(X)$, and the
corresponding probability density $\rho(x)={\rm d}P(x)/{\rm d} x$:
\begin{equation}
\label{eq: 0} \rho^\ast(x\vert N)\sim N^3\rho(x) \biggr[\int_0^x\!
{\rm d}x_0\, \rho(x_0)P(x_0)\biggl]\exp[-N P(x)]\ .
\end{equation}
(We use $A(N)\sim B(N)$ to mean that $A(N)$ and $B(N)$ are
asymptotically equal in the limit as $N\to \infty$.) For
comparison, the corresponding probability density for the strength
of a randomly assembled chain is
\begin{equation}
\label{eq: -1} \rho(x\vert N)\sim N\rho(x)\exp[-NP(x)]\ .
\end{equation}
Figure \ref{fig: 1} shows the comparison between equation
(\ref{eq: 0}) and a histogram of data from a numerical simulation
of the chain-breaking process, for a particular choice of link
strength distribution $\rho(x)$.

We were motivated to study this chain-breaking process because it
models fragmentation process in which the distribution of
strengths of fragments (rather than the distribution of fragment
sizes) is the principal concern. We are not aware of any earlier
investigations of the strength distribution of fragments. The
chain-breaking process is of interest because it is a fundamental
model for such processes, which has an explicit solution for the
distribution $\rho^\ast(x\vert N)$. The model also has direct
application to the degradation of long chain polymers.

In section 2 we consider the distribution of the sizes of
fragments of a repeatedly broken chain, and section 3 describes
elementary results for randomly assembled chains. In section 4
these expressions are used to obtain a recursion for a sequence of
distributions which are related to $\rho^\ast(x\vert N)$. Finally
section 5 shows how the limiting distribution of this sequence is
obtained self-consistently, and used to derive equation (\ref{eq:
0}) above.

\section{Distribution of fragment sizes}
\label{sec: 2}

First we discuss the distribution of sizes of chain fragments.
After $i$ steps of splitting the chain, we have $2^i$ fragments.
Let $W_i(N)$ be the number of segments of lengths $N$ at step $i$,
and let us consider the case where these numbers are so large that
it is sufficient to calculate expectation values and ignore the
statistical fluctuations of $W_i(N)$ (this assumption is valid in
the limit as ${\cal N}\to \infty$). Using the fact that the
position of the weakest link has equal probability to be at any
site, these numbers satisfy a recursion relation
\begin{equation}
\label{eq: 1} W_{i+1}(N)=\sum_{M=N+1}^\infty {2\over{M-1}}W_i(M)\ .
\end{equation}
Rather than following this iteration for a single chain being
broken, it is easier to consider a steady state $W(N)$, with
destruction of one additional chain being initiated at each step.
Thus we seek to solve
\begin{equation}
\label{eq: 2} W(N)=\sum_{M=N+1}^\infty {2\over{M-1}}W(M)\ .
\end{equation}
We find $W(N)\sim C/N^2$ for $N\to \infty$ for some constant $C$.

\section{Strength of a random chain segment}
\label{sec: 3}

By way of preparation we discuss elementary results on the
distribution of strengths for a chain with completely random
links. Let $P(x\vert N)$ be the probability that a chain of length
$N$ breaks at a tension which is less than $x$. For $N=1$, we have
$P(x\vert 1)=P(x)$. The probability that a chain of $N$ links is
unbreakable at tension $x$ is $1-P(x\vert N)=[1-P(x)]^N$. In the
limit where $N\gg 1$, the power may be approximated by
exponentiation: we find $P(x\vert N)\sim 1-\exp[-NP(x)]$ (which
gives (\ref{eq: -1}) upon differentiation).

We also require the conditional probability $P_{\rm c}(x\vert
x_0,N)$ that a chain of length $N$ breaks at tension $x$ if we
know that it is definitely not broken by a tension $x_0$. In this
case we know that the probability of the strength of an individual
element being less than $x$ is $[P(x)-P(x_0)]/[1-P(x_0)]$, so that
\begin{equation}
\label{eq: 5} P_{\rm c}(x\vert
x_0,N)=1-\left[\frac{P(x)-P(x_0)}{1-P(x_0)}\right]^N
\end{equation}
provided that $x>x_0$ (and zero otherwise). When $N\gg 1$, this
may be approximated by
\begin{equation}
\label{eq: 6} P_{\rm c}(x\vert x_0,N)\sim\bigl[
1-\exp\bigl(-N[P(x)-P(x_0)]\bigr)\bigr]\Theta (x-x_0)
\end{equation}
where $\Theta (x)$ is the unit increasing step function (Heaviside
function).

\section{A recursion relation for probabilities of successive breaking tensions}
\label{sec: 4}

Our objective is to obtain the probability density
$\rho^\ast(x\vert N)$ that a chain of length $N$, which has been
produced by repeatedly breaking a very long chain, has breaking
strength $x$.

We determine this distribution of strengths by calculating a
related distribution. Consider the subdivision of the chain,
starting from a very long chain of length ${\cal N}$. When this is
repeatedly split, for convenience we always discuss the leftmost
segment. At the $i$th stage of subdivision, this segment having
length $N_i$ is produced by breaking a link of strength $x_i$. Let
the probability density for the strength of this broken link be
$\rho_i(x_i\vert N_i)$. The corresponding probability for the
$i$th split to occur at a tension less than $x_i$ for a segment of
length $N_i$ is $P(x_i\vert N_i)$. We shall obtain a recursion
formula which expresses $\rho_{i+1}(x_{i+1},N_{i+1})$ in terms of
this distribution. The chain fragment at stage $i+1$ is produced
by breaking a segment of length $N_i$, in which all of the links
are known to be stronger than $x_i$. Both of these items of prior
information have uncertain values.

Consider first the distribution of values of $N_i$, for a given
value of $N_{i+1}$. We have already seen (equation (\ref{eq: 2})) that repeated random
sub-division of an interval produces a steady-state distribution
of lengths $W(N)\sim 1/N^2$. Subdivision
of an interval of length $N_i>N_{i+1}$ produces an interval of
length $N_{i+1}$ with probability $1/N_i$. The probability
distribution for $N_i$ is therefore proportional to $N_i^{-3}$ for
$N_i>N_{i+1}$ (and zero otherwise). For $N_{i+1}\gg 1$, the
normalised distribution of $N_i$ is therefore
\begin{equation}
\label{eq: 7} P_{\rm n}(N_i\vert N_{i+1})\sim \left\{
\begin{array}{ll} {2N_{i+1}^2\over{N_i^3}}&\ \ \ N_{i+1}<N_i\cr
0&\ \ \ N_{i+1}\ge N_i
\end{array}\right .
\ .
\end{equation}
The probability $P_{i+1}(x_{i+1}\vert N_{i+1})$ is obtained from
$P_{\rm c}(x_{i+1}\vert x_i,N_i)$ by averaging over the
probability densities of both $N_i$ and $x_i$. (Note that the
latter probability density is related to the unknown function that
we wish to calculate.):
\begin{equation}
\label{eq: 8} P_{i+1}(x_{i+1}\vert
N_{i+1})=\sum_{N_i=N_{i+1}+1}^\infty P_{\rm n}(N_i\vert N_{i+1})
\int_0^{x_{i+1}} {\rm d}x_i\ P_{\rm c}(x_{i+1}\vert x_i,N_i)
\rho_i (x_i\vert N_i)\ .
\end{equation}
Differentiating (\ref{eq: 8}) with respect to $x_{i+1}$ and
substituting the known expressions for $P_{\rm n}(N_i\vert
N_{i+1})$ and $P_{\rm c}(x_1\vert x_0,N_0)$ we obtain
\begin{equation}
\label{eq: 9} \rho_{i+1}(x_{i+1}\vert N_{i+1})\sim
\sum_{N_i=N_{i+1}+1}^\infty\frac{2N_{i+1}^2}{N_i^2}\int_0^{x_{i+1}}
{\rm d}x_i\ \rho(x_i)
\exp\bigl(-N_i[P(x_{i+1})-P(x_i)]\bigr)\rho_i (x_i\vert N_i)\ .
\end{equation}

\section{Self-consistent solution}
\label{sec: 5}

We assume that the distribution $\rho_i(x\vert N)$ becomes
independent of the generation index $i$ when we consider the
subdivision of a very long chain. This leads to a \lq
self-consistency' condition for the probability density
$\rho_i(x\vert N)={\rm d}P_i(x\vert N)/{\rm d}x$. Approximating
the sum in equation (\ref{eq: 9}) by an integral, and replacing
$\rho_i$ and $\rho_{i+1}$ by the asymptotic, self-consistent
function $\rho_\infty$ yields
\begin{equation}
\label{eq: 10} \rho_\infty (x\vert N)\sim 2N^2 \int_N^\infty {\rm
d}N_0\ {1\over{N_0^2}}\int_0^x {\rm d}x_0\
\exp\bigl(-N_0[P(x)-P(x_0)]\bigr)\rho_\infty (x_0\vert N_0)\ .
\end{equation}
We assume that the corresponding probability $P_\infty(x\vert N)$
is of the form $P_\infty (x\vert N)=F(NP(x))$ for some function
$F$ (which increases monotonically from $F(0)=0$ to
$F(\infty)=1$). Writing $w=P(x)$, the derivative $f=F'$ of $F$  satisfies
\begin{equation}
\label{eq: 11} f(Nw)=2N\int_N^\infty {\rm d}N_0 \
{1\over{N_0}}\int_0^w {\rm d}w_0\ \exp[-N_0(w-w_0)] f(N_0w_0)\ .
\end{equation}
We express this integral equation in terms of $g(X)=\exp(X)f(X)$,
and differentiate. We find that the primitive of $g(X)$, namely
$G(X)$, satisfies
\begin{equation}
\label{eq: 15} G''(X)=\frac{X+1}{X}G'(X)-{2\over{X}}G(X)\ .
\end{equation}
The solution of this equation is of the form
\begin{equation}
\label{eq: 16} G(X)=AX^2+B[\exp(-X)(1+X)-X^2{\rm Ei}(X)]
\end{equation}
where $A$, $B$ are constants and ${\rm Ei}(X) = {\cal
P}\int_{-\infty}^X\!\! {\rm d}z \, {\rm e}^{z}/z$ is the
exponential integral. The requirement that $G(0)=0$ implies that
$B=0$, so that the normalised function $f(x)$ is $f(x)=x\exp(-x)$.
Thus we find the probability density for a chain segment of length
$N$ being formed by breaking a link of strength $x$ in the form
\begin{equation}
\label{eq: 18} \rho_\infty(x\vert N)=N^2\rho(x)P(x)\exp[-NP(x)]\ .
\end{equation}

Finally, we are in the position to obtain
the desired result,
the probability density for a segment of
length $N$ having a weakest link of strength $x$. This is obtained
from the distribution $\rho_\infty(x\vert N)$ as follows
\begin{eqnarray}
\label{eq: 19} \rho^\ast(x\vert N)&=&\int_0^x{\rm d}x_0\
\frac{{\rm d}}{{\rm d}x}P_{\rm c}(x\vert x_0,N)\rho_\infty(x_0\vert N)\nonumber \\
&=& N\rho(x) \int_0^x {\rm d}x_0\
\exp\big(-N[P(x)-P(x_0)]\bigr)\rho_\infty (x_0\vert N)\ .
\end{eqnarray}
In the limit of large $N$ we find the simple, asymptotically exact
result (\ref{eq: 0}).

In the case where $\rho(x)$ has a finite limit for very weak
chains, $\rho_0=\rho(0)$, the strength distribution (\ref{eq: 0})
has a simple universal scaling form when $N\gg 1$:
\begin{equation}
\label{eq: 21} \rho^\ast(x\vert N)\sim
{\textstyle{\frac{1}{2}}}N^3\rho_0^3x^2\exp(-N\rho_0 x)\ .
\end{equation}

\begin{acknowledgements}
BM acknowledges support from Vetenskapsr\aa{}det.
\end{acknowledgements}

%
%

\end{document}